\documentclass[aps,twocolumn,superscriptaddress,floatfix,prl]{revtex4}
\usepackage{graphicx,subfigure,amsmath}
\begin{document}
\title{Quantum Tunneling Detection of Two-photon and Two-electron Processes}
\date{\today}

\author{J. Tobiska}
\author{J. Danon}
\affiliation{Kavli Institute of NanoScience, Delft University of Technology, 2628 CJ Delft, The Netherlands}
\author{I. Snyman}
\affiliation{Instituut-Lorentz, Universiteit Leiden, P.O. Box 9506, 2300 RA Leiden, The Netherlands}
\author{Yu.\ V. Nazarov}
\affiliation{Kavli Institute of NanoScience, Delft University of Technology, 2628 CJ Delft, The Netherlands}
\begin{abstract}
    We analyze the operation of a quantum tunneling detector coupled to a coherent conductor. We demonstrate that in a certain energy range the output of the detector is determined by two-photon processes, two-electron processes
and the interference of the two. We show how the individual contributions of these processes can be resolved in experiments.
\end{abstract}
\maketitle
The quantum nature of electron transfer in coherent conductors is seldom explicitly manifested in averaged current-voltage curves. To reveal it one should measure current noise and/or higher-order correlations of current comprising Full Counting Statistics which arise from the transfer~\cite{shotnoise}. Such measurements not only reveal the discrete nature of the charges transferred, they also quantify quantum many-body effects in electron transport and may be used for the detection of pairwise entanglement of transferred particles~\cite{qnoise,antoniospinent,beenakker:056801,samuelsson:026805}.
If the noise is measured at frequencies in the quantum range, $\hbar\omega\gg k_\mathrm{B}T$, the measurement amounts to the detection of photons produced by the current fluctuations. This aspect is important in view of attempts to transfer quantum information from electrons to photons and back~\cite{elphtransfer}.

It was demonstrated in~\cite{noisedetection} that one needs a quantum
detector to measure quantum noise. Indeed, any classical measurement of
a fluctuating quantity would give a noise spectrum symmetric in
frequency, $S(\omega)=S(-\omega)$. A {\em quantum tunneling detector} is
generally a quantum two-level system with a level separation
$\varepsilon>0$. The result of detection are two transition rates:
$\Gamma_\mathrm{up}$ from the lower to the higher level and
$\Gamma_\mathrm{down}$ for the reverse direction. The most probable
transitions are accompanied by either absorption or emission of a photon
of matching energy $\hbar\omega=\varepsilon$. One can define the noise
spectrum in such a way that it is proportional to the transition rates
$S(\pm\varepsilon/\hbar)\propto \Gamma_\mathrm{up,down}(\varepsilon)$.
Differences between $\Gamma_\mathrm{up,down}$ thus manifest the quantum
nature of noise. If the source of noise is a coherent conductor biased
by a voltage $V$, detector signals in the range $\varepsilon<eV$ are
readily interpreted in terms of single electron transfers through the
conductor. The maximum energy gain available for electrons in the course
of such transfer is $eV$. Consequently this value also limits the energy
of the emitted photon.

A first proposal for the experimental realization of a quantum tunneling detector included transitions between two localized electron states in a double quantum dot~\cite{dqdnoisedetector}. However it does not matter much if the tunneling occurs between localized or delocalized electron states and if all tunnel events are accompanied by the same energy transfer $\varepsilon$. In most practical cases the energy dependence of the rates $\Gamma_\mathrm{up,down}$ can be readily extracted from the measurement results. This is why quantum tunneling detection has been experimentally realized in a superconducting tunnel junction~\cite{qnoisetwolevel} and in a single quantum dot~\cite{dotpat}.

In this Letter we study quantum tunneling detection in the range
$eV<\varepsilon<2eV$ assuming $\varepsilon,eV\gg k_\mathrm{B}T$. The
motivation is that for these energies the detector is not sensitive
to single-electron one-photon processes described above and its output
--- the transition rate $\Gamma_\mathrm{up}$--- is determined by much more
interesting two-particle processes. It is clear from plain energy
considerations that transitions may originate from {\em two-photon}
processes. Such two-photon absorption can occur given any
non-equilibrium photon distribution bounded by $eV$, not necessarily
produced by a coherent conductor. Less obvious and specific for a
coherent conductor is a cooperative {\em two-electron} process. Indeed,
if two electrons team up in crossing the conductor they can emit a
single photon with an energy up to $2eV$. Essential for this cooperation
are electron-electron interactions. It is known~\cite{IngoldNazarov} that
the most important electron-electron interaction in this energy range is
due to the electromagnetic environment of the conductor, the same
environment in which the non-equilibrium photons dwell.
\begin{figure}[b]
\includegraphics[width=7cm]{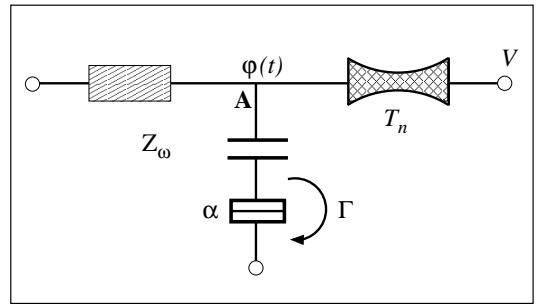} 
\caption{Model circuit}\label{fig:system}
\end{figure}
We quantify the signals due to
two-photon and two-electron events and find them to be of the same order
of magnitude. We also show that part of the signal is due to quantum
interference of these two processes: {\it one-and-half-photon} absorption
events. We demonstrate how different contributions can be separated in
experiments thereby facilitating the direct observation of two-particle
processes in the context of quantum transport.

We concentrate on a model circuit consisting of four elements as given
in figure~\ref{fig:system}. A voltage biased coherent contact
characterized by a set of transmission eigenvalues $\{T_n\}$ is embedded
in an electromagnetic environment with impedance $Z_\omega$. The
environment transforms the current fluctuations in the conductor to
voltage fluctuations in node A
which are conveniently expressed in terms of a
phase $\varphi=\frac{e}{\hbar}\int dt V(t)$. The most general model
including the detector and coherent contact would be a
four-pole circuit studied in~\cite{kindermann:035336} that couples
two poles of the detector with two poles of the contact.
Here, we concentrate on
the experimentally relevant case of
capacitive coupling. Owing to voltage division between two capacitors,
the detector senses a fraction
$0<\alpha<1$ of the voltage fluctuations in node A. We will see that
changing the ``visibility'' parameter $\alpha$, enables the separation of
two-electron and two-photon processes in experiments. The relevant impedance
is made up of an environmental impedance combined with that of two capacitors and that of the coherent contact. We measure this impedance $z_\omega$ in units of $R_K \equiv 2\pi \hbar/e^2$ and assume the low-impedance limit
$z_\omega \ll 1$; this provides us with a physically justified
small parameter.

The detector consists of two
localized charge states connected by a tunnel amplitude ${\cal
T}$. In the presence of voltage fluctuations in the node A,
the amplitude is modified as follows: ${\cal T}(t)\to {\cal T}e^{i\alpha\varphi(t)}$.
In perturbation theory, the inelastic tunneling rate between two states
separated by $\varepsilon$ is given by correlators of
$\alpha{\varphi}(t)$~\cite{IngoldNazarov}
\begin{equation}
    \Gamma(\varepsilon)=\frac{|{\cal T}|^2}{2\pi\hbar^2}\int dt\langle
e^{i\alpha{\varphi}(t)}e^{-i\alpha{\varphi}(0)}\rangle
e^{\frac{i}{\hbar}\varepsilon t}.\label{eq:pofe}
\end{equation}
The rate $\Gamma(\varepsilon)$ is
therefore the Fourier transform
of the correlation function
$\langle e^{i\alpha{\varphi}(t)}e^{-i\alpha{\varphi}(0)}\rangle$,
 $\alpha{\varphi}(t)$ being the phase fluctuations over the
detector. From now on we take $\hbar=e=k_\mathrm{B}=1$.

Equation (\ref{eq:pofe}) tells us that the inelastic tunneling rates in the detector are completely determined by the voltage fluctuations over the
junction. Therefore measuring the inelastic current through the dots
we are sensitive to the noise spectrum of the environment.

To evaluate $\langle
e^{i\alpha{\varphi}(t)}e^{-i\alpha{\varphi}(0)}\rangle$
we construct a path integral representation of this quantity using a
non-equilibrium Keldysh technique \cite{RammerSmith}
for quantum-circuits~\cite{kindermcurvoltbias}
\begin{equation}
\begin{split}
\langle
e^{i\alpha{\varphi}(t)}e&^{-i\alpha{\varphi}(0)}\rangle=
\int\mathcal{D}[\boldsymbol\phi]\exp\{-iS_{\mathrm{env}}[\boldsymbol\phi]\\
&-iS_{\mathrm{cond}}[\boldsymbol\phi]+i\alpha[-\varphi^+(0)+\varphi^-(t)]\}.
\end{split}
\label{eq:pathint}
\end{equation}
The integration goes over the time-dependent
fluctuating fields $\varphi^\pm(t)$ in node $A$,
$\pm$ corresponding to the forward (backward) part of the
Keldysh contour.
$S_{\mathrm{env}}$ and $S_{\mathrm{cond}}$ are
the contributions to the
Keldysh action originating from the environment and the coherent conductor
respectively.

Since the environment is linear, its action is quadratic in the fields and
at zero temperature reads (cf.~\cite{KindermannNazarov2})
\begin{equation}
S_{\mathrm{env}}=\int d\omega \, \boldsymbol\phi_{-\omega}^{\,T}
A(\omega) \boldsymbol\phi_{\omega}
\end{equation}
with
\begin{equation*}
A(\omega)=-\frac{i}{2}\left({\begin{array}{*{20}c}
0 & -\frac{\omega}{z_{-\omega}}\\
\frac{\omega}{z_\omega} & |\omega|\mathrm{Re}\{\frac{1}{z_\omega}\}\\
\end{array}}\right),
\end{equation*}
$z_\omega$ being the corresponding impedance.
We use Fourier transformed
fields $\boldsymbol\phi_{\omega}=\left(\phi_{\omega}, \chi_{\omega}
\right)^T$ defined with $\varphi^\pm=\phi \pm \frac{1}{2}\chi$.

All non-quadratic contributions to the action
originate from the coherent conductor. The action
$S_{\mathrm{cond}}$ can be expressed in terms of Keldysh Green
functions $\check{G}_{L,R}$ of electrons in the reservoirs left and
right of the contact \cite{KindermannNazarov2}
\begin{equation}
S_{\mathrm{cond}}=\frac{i}{2} \sum_n \mathrm{Tr} \ln
[1+\frac{T_n}{4}(\{\check{G}_L(\boldsymbol\phi),\check{G}_R\}-2)].
\label{eq:sqpc}
\end{equation}
The fields $\boldsymbol\phi$ enter in this action
via the gauge transform of $\check{G}_L$.\cite{KindermannNazarov2}

To comprehend the physics involved, let us first disregard any non-quadratic parts
and take only the quadratic part of $S_\mathrm{cond}$.
In this case the path integral is Gaussian,
and can be evaluated exactly. We recover the well known result from $P(E)$-theory
(cf.~\cite{dqdnoisedetector,IngoldNazarov}): $\langle
e^{i\alpha{\varphi}(t)}e^{-i\alpha{\varphi}(0)}\rangle=\exp[J(t)]$
with
\begin{equation}\label{eq:jt}
J(t)=\langle \alpha\varphi(t)\alpha\varphi(0)\rangle=\alpha^2\int
d\omega\frac{|z_{\omega}|^2}{\omega^2}K(\omega)[e^{-i\omega
t}-1].
\end{equation}
The impedance includes
the dimensionless conductance $g_c \equiv \sum_n T_n$
of the contact.

In the limit of $T=0$ we find in agreement with
results of~\cite{noisedetection,dqdnoisedetector}
\begin{equation}
\begin{split}
    K(\omega)=g_c &\{FD(\omega+V)+(2-2F)D(\omega)\\
&+FD(\omega-V)\}+2\mathrm{Re}\{\frac{1}{z_\omega}\}D(\omega)
\end{split}
\end{equation}
with
$D(\omega)\equiv-\omega\theta(-\omega)$ and the Fano factor
$F \equiv \sum_n T_n(1-T_n) / \sum_n T_n$.
The first term in $K(\omega)$ ($\propto g_c$) represents
the non-equilibrium
current noise spectrum of the coherent contact that vanishes  for $\omega>V$.
In physical terms this means
that the highest energy $\omega$
an electron can emit traversing the conductor is exactly $V$.
The second part represents the spectrum of the environment.
It is zero for $\omega>0$,
since the environment can only absorb energy at $T=0$.

The time-dependent part of $J(t)$ is the Fourier transform of
$K(\omega)|z_\omega|^2/\omega^2$ and $\Gamma(\varepsilon)$ is in turn the
Fourier transform of $\exp[J(t)]$. If we expand $\exp[J(t)]$ in terms of
$J(t)$, the $n$-th term presents the contribution
of a process involving absorption of $n$ photons in the detector.
Such an $n$-photon process dominates
in the interval $(n-1)V<\varepsilon<nV$ and its contribution
is proportional to $\alpha^{2n}$.

\begin{figure}
\includegraphics[width=7cm]{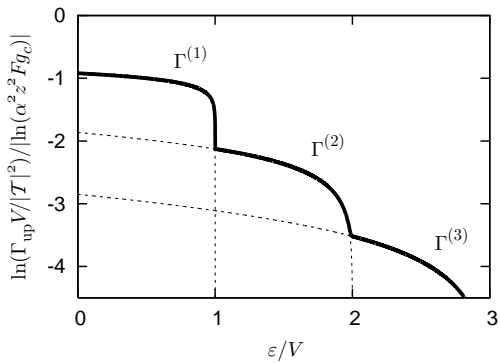}
\caption{$n$-photon contributions (dashed) to the detector output. Each contribution dominates in the energy interval $(n-1)V<\varepsilon<nV$. Subsequent contributions are suppressed by a factor $\approx \alpha^2 z^2 g_c F $. This is seen as a staircase structure in the log plot. The sum of all contributions is given by the solid line. To produce this plot, we took $z_\omega/\omega = z/V$, $z=0.01$, $g_c F =1.75$, $\alpha=1$.}
\label{fig:pelogplot}
\end{figure}

The one-photon contribution gives $\Gamma^{(1)}_\mathrm{up}/\Gamma_\mathrm{down}
\simeq  z g_c F$. Each extra photon brings in a small factor, such that $\Gamma^{(n+1)}/\Gamma^{(n)} \simeq \alpha^2 z^2 g_c F$.
This is seen as a staircase in the log plot
presented in Figure \ref{fig:pelogplot}.

What we did was wrong since we did not take into account the non-quadratic
terms in the action. These describe more interesting  many-electron
processes and are---as we show
below---of the same order of magnitude.
Since the path integral in (\ref{eq:pathint}) can not be evaluated in general,
we proceed by perturbative expansion.

\begin{figure}
    \includegraphics[width=6.5cm]{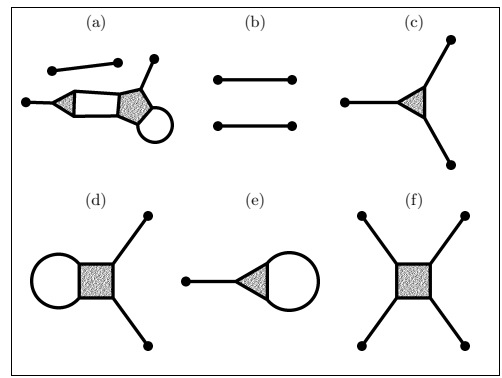}
    \caption{Typical Feynman diagrams
    appearing in the expansion
    of the path integral representation of the tunneling rate.}
    \label{fig:feynm}
\end{figure}

Indeed, since $|z|\ll 1$, the Gaussian part of the action, being proportional
to $z^{-1}$, suppresses fluctuations in the path integral and we can
treat the remaining part perturbatively. First we expand
$-iS_\mathrm{cond}[\boldsymbol\phi]$ around $\boldsymbol\phi=0$. As
mentioned previously (see the discussion below equation (\ref{eq:jt})),
the first and second order terms just renormalize the impedance. The
exponential of the remaining higher order terms is then again expanded
in $\boldsymbol\phi$ around $\boldsymbol\phi=0$. This expansion may be
represented in terms of diagrams such as those in
figure~\ref{fig:feynm}. Diagram (a) represents a high order term, from
which the general structure becomes clear: Diagrams consist of lines,
polygons and external vertices. The expansion contains not only
connected diagrams, but all disconnected diagrams as well. A polygon
with $n$ vertices is associated with the symmetrized $n$-th order
coefficient in the Taylor expansion of
$-iS_\mathrm{cond}[\boldsymbol\phi]$. Each polygon contributes a factor $g_c$.
Lines represent propagators of
$\boldsymbol\phi$ corresponding to the Gaussian action with renormalized
impedance. They are of order $z$ making $n$-line diagrams $z^n$ in
leading order. External vertices (indicated by dots in the figure) are
associated with the time-dependent linear term
$i\alpha[-\varphi^+(0)+\varphi^-(t)]$ in equation (\ref{eq:pathint}).
Thus a diagram with $s$ external vertices gives a correction
proportional to $\alpha^s$. Furthermore, diagrams without external
vertices are time-independent and according to equation (\ref{eq:pofe})
contribute only to elastic tunneling processes. Diagrams (b) to (f)
represent some of the lowest order terms in the expansion.

We consider transitions in the detector where energies between $V$ and
$2V$ are absorbed. In this interval, the detector output
is given by diagrams (b), (c) and (d), which are
proportional to $\alpha^4,\alpha^3$ and $\alpha^2$ respectively.

From the results presented in figure \ref{fig:pelogplot} we have learned that $n$-photon processes come with a coefficient $\alpha^{2n}$. Hence the $\alpha^3$ contribution is not readily expected:
it seems to signal a process with {\it one-and-half}
photons absorbed.
We disregard diagram (e) which only contributes to elastic processes. In
the energy interval considered, the combined $z^3$ contribution of the
included diagrams is zero and we obtain a tunneling rate that goes as
$g^2_c z^4$. Since a diagram like (f) has four lines,
it could potentially contribute to the current with the same order in $z$. However, its contribution can only be proportional to $g_c$ and is disregarded.

The expansion of $S_\mathrm{cond}$ up to fourth order terms and
subsequent evaluation of the diagrams is straightforward but requires
rather involved and lengthy calculations.
Fortunately in the interval of
interest the three contributions can be
combined in a compact expression
\begin{eqnarray}
{\Gamma}_\mathrm{up}= 2 |{\cal T}|^2 g^2_c F^2 \int_{\varepsilon-V}^V
d\omega(V-\omega)(\varepsilon-V-\omega) \nonumber \\
\frac{|z_\omega|^2}{\omega^2}\left|\frac{\alpha^2}{2}
\frac{z_{\varepsilon-\omega}}{\varepsilon-\omega}+
\alpha\frac{z_\varepsilon}{\varepsilon}\right|^2 ,\label{eq:gammaiz}
\end{eqnarray}
which is the main result of
our work. The rate is proportional to the
square of the zero-frequency current noise
$S_\mathrm{cond}(0)=\frac{2}{\pi}g_c F$.

The part proportional to $\alpha^4$ (diagram b)
represents a two-photon process originating from the quadratic part of $S_\mathrm{cond}$ and was already present in figure \ref{fig:pelogplot}.
We have thus shown that there are contributions of
the same order resulting from non-linearities in the conductor.
The $\alpha^2$ term (diagram d) is the result of the
two-electron and one-photon process
expected from general reasoning presented in the introduction.
We see that the $\alpha^3$ term
comes from the cross-term in the modulus square.
This unambiguously
identifies digram (c) as the result of
quantum interference of the two-electron process and the two-photon processes---
an interpretation that was not obvious from the beginning.

To understand this interference, we note that the photon modes
involved are delocalized across the whole circuit. A photon in each mode
can be absorbed in the detector as well as in the environment or the
contact. An elementary process is such that the final state
differs from the initial state by two photons absorbed in two given modes.
The final state can be reached by two amplitudes:
one with both photons absorbed in the detector and
one with a photon absorbed in the detector and a photon absorbed
in the environment.
While the squares of these amplitudes represent the
probabilities of two-photon and two-electron processes respectively,
their cross-term gives rise to an interference contribution $\propto \alpha^3$.

\begin{figure}
\includegraphics[width=\columnwidth]{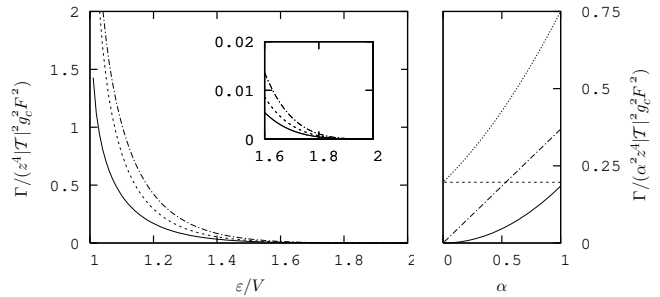}
\caption{Left: Contributions to the detector output
due to two-photon (solid), two-electron (dashed) processes and
their interference (dash-dotted) at $\alpha=0.8$ versus detector
level splitting. The inset
presents a zoom at $\varepsilon\to 2V$. Right: Different dependence
on the coupling strength $\alpha$ enables experimental
identification of these three contributions. The contributions
and their sum (dotted) are plotted for $\varepsilon=1.3V$. }
\label{fig:comp}
\end{figure}

The simplest concrete model is that of a frequency-independent
impedance, $z_\omega = z$ at $\omega \simeq V$.
The integration in Eq. \ref{eq:gammaiz} yields
for the three distinct contributions
( $\tilde{\varepsilon}=\varepsilon/V$, $1<\tilde{\varepsilon}<2$ )
\begin{equation}
	\frac{\Gamma_i}{z^4|{\cal T}|^2g_c^2 F^2}= \left\{
    \begin{array}{l}
\alpha^4\left[-\frac{2-2\tilde{\varepsilon}+\tilde{\varepsilon}^2}{\tilde{\varepsilon}^3}\ln(\tilde{\varepsilon}-1)-\frac{2-\tilde{\varepsilon}}{\tilde{\varepsilon}^2}\right]\\
        2\alpha^3 \left[-\frac{2-2\tilde{\varepsilon}+\tilde{\varepsilon}^2}{\tilde{\varepsilon}^3}\ln(\tilde{\varepsilon}-1)-\frac{2-\tilde{\varepsilon}}{\tilde{\varepsilon}^2}\right]\\      \alpha^2\left[-\frac{2}{\tilde{\varepsilon}}\ln(\tilde{\varepsilon}-1)-\frac{4(2-\tilde{\varepsilon})}{\tilde{\varepsilon}^2}\right]
    \end{array}
    \right.
    \label{eq:rate123}
\end{equation}
All contributions scale as $(\varepsilon -2V)^3$ at the two-photon threshold
and logarithmically diverge at approaching the one-photon threshold.
(see Fig. \ref{fig:comp}).

Here, we have quantified the contributions for a very specific non-linear
quantum noise source: a coherent conductor.
However in the case of any unknown source of this kind the $\alpha$-dependence of the contributions allows one to separate and identify them experimentally 
(right pane of Fig. \ref{fig:comp}).
One would measure the detector output changing the coupling to the detector.
Formally, three measurements at three different $\alpha$ are sufficient
to determine the relative strength of all three contributions.
In any case, in the limit of small coupling $\alpha \to 0$ the
detector output is dominated by two-electron events.
Further characterization may be achieved by engineering of a
frequency-dependent impedance. For instance, setting $z(\omega = \epsilon)$ to 0
kills both interference and two-electron contribution.

To conclude, we have shown that the quantum tunneling detector in
the energy interval specified is selectively sensitive to two-particle
processes.
The detector output is generally determined by three contributions:
two-photon processes, two-electron
processes and the interference of the two. These three sources can be distinguished
experimentally by measuring at different couplings $\alpha$ to the detector.
Our results thus facilitate the direct observation of many-particle events
in the context of quantum transport. This work was supported by the Dutch Foundation for Fundamental Research on Matter (FOM). 
\bibliography{qdqpc}

\begin{thebibliography}{15}
\expandafter\ifx\csname natexlab\endcsname\relax\def\natexlab#1{#1}\fi
\expandafter\ifx\csname bibnamefont\endcsname\relax
  \def\bibnamefont#1{#1}\fi
\expandafter\ifx\csname bibfnamefont\endcsname\relax
  \def\bibfnamefont#1{#1}\fi
\expandafter\ifx\csname citenamefont\endcsname\relax
  \def\citenamefont#1{#1}\fi
\expandafter\ifx\csname url\endcsname\relax
  \def\url#1{\texttt{#1}}\fi
\expandafter\ifx\csname urlprefix\endcsname\relax\def\urlprefix{URL }\fi
\providecommand{\bibinfo}[2]{#2}
\providecommand{\eprint}[2][]{\url{#2}}

\bibitem[{\citenamefont{Blanter and B{\"u}ttiker}(2000)}]{shotnoise}
\bibinfo{author}{\bibfnamefont{Y.~M.} \bibnamefont{Blanter}} \bibnamefont{and}
  \bibinfo{author}{\bibfnamefont{M.}~\bibnamefont{B{\"u}ttiker}},
  \bibinfo{journal}{Phys. Rep.} \textbf{\bibinfo{volume}{336}},
  \bibinfo{pages}{1} (\bibinfo{year}{2000}).

\bibitem[{\citenamefont{Nazarov}(2003)}]{qnoise}
\bibinfo{editor}{\bibfnamefont{Y.~V.} \bibnamefont{Nazarov}}, ed.,
  \emph{\bibinfo{title}{Noise in Mesoscopic Physics}},
  vol.~\bibinfo{volume}{97} of \emph{\bibinfo{series}{NATO Science Series}}
  (\bibinfo{publisher}{Kluwer Academic Publishers}, \bibinfo{address}{NATO
  Advanced Research Workshop on Quantum Noise in Mesoscopic Physics, Delft, The
  Netherlands}, \bibinfo{year}{2003}).

\bibitem[{\citenamefont{Lorenzo and Nazarov}(2005)}]{antoniospinent}
\bibinfo{author}{\bibfnamefont{A.~D.} \bibnamefont{Lorenzo}} \bibnamefont{and}
  \bibinfo{author}{\bibfnamefont{Y.~V.} \bibnamefont{Nazarov}},
  \bibinfo{journal}{Phys. Rev. Lett.} \textbf{\bibinfo{volume}{94}},
  \bibinfo{eid}{210601} (\bibinfo{year}{2005}).

\bibitem[{\citenamefont{Beenakker and Kindermann}(2004)}]{beenakker:056801}
\bibinfo{author}{\bibfnamefont{C.~W.~J.} \bibnamefont{Beenakker}}
  \bibnamefont{and}
  \bibinfo{author}{\bibfnamefont{M.}~\bibnamefont{Kindermann}},
  \bibinfo{journal}{Phys. Rev. Lett.} \textbf{\bibinfo{volume}{92}},
  \bibinfo{eid}{056801} (\bibinfo{year}{2004}).

\bibitem[{\citenamefont{Samuelsson et~al.}(2004)\citenamefont{Samuelsson,
  Sukhorukov, and B{\"u}ttiker}}]{samuelsson:026805}
\bibinfo{author}{\bibfnamefont{P.}~\bibnamefont{Samuelsson}},
  \bibinfo{author}{\bibfnamefont{E.~V.} \bibnamefont{Sukhorukov}},
  \bibnamefont{and}
  \bibinfo{author}{\bibfnamefont{M.}~\bibnamefont{B{\"u}ttiker}},
  \bibinfo{journal}{Phys. Rev. Lett.} \textbf{\bibinfo{volume}{92}},
  \bibinfo{eid}{026805} (\bibinfo{year}{2004}).

\bibitem[{\citenamefont{Cerletti et~al.}(2004)\citenamefont{Cerletti, Gywat,
  and Loss}}]{elphtransfer}
\bibinfo{author}{\bibfnamefont{V.}~\bibnamefont{Cerletti}},
  \bibinfo{author}{\bibfnamefont{O.}~\bibnamefont{Gywat}}, \bibnamefont{and}
  \bibinfo{author}{\bibfnamefont{D.}~\bibnamefont{Loss}},
  \bibinfo{journal}{cond-mat/0411235}  (\bibinfo{year}{2004}).

\bibitem[{\citenamefont{Gavish et~al.}(2000)\citenamefont{Gavish, Levinson, and
  Imry}}]{noisedetection}
\bibinfo{author}{\bibfnamefont{U.}~\bibnamefont{Gavish}},
  \bibinfo{author}{\bibfnamefont{Y.}~\bibnamefont{Levinson}}, \bibnamefont{and}
  \bibinfo{author}{\bibfnamefont{Y.}~\bibnamefont{Imry}},
  \bibinfo{journal}{Phys. Rev. B} \textbf{\bibinfo{volume}{62}},
  \bibinfo{pages}{R10637} (\bibinfo{year}{2000}).

\bibitem[{\citenamefont{Aguado and Kouwenhoven}(2000)}]{dqdnoisedetector}
\bibinfo{author}{\bibfnamefont{R.}~\bibnamefont{Aguado}} \bibnamefont{and}
  \bibinfo{author}{\bibfnamefont{L.~P.} \bibnamefont{Kouwenhoven}},
  \bibinfo{journal}{Phys. Rev. Lett.} \textbf{\bibinfo{volume}{84}},
  \bibinfo{pages}{1986} (\bibinfo{year}{2000}).

\bibitem[{\citenamefont{Deblock et~al.}(2003)\citenamefont{Deblock, Onac,
  Gurevich, and Kouwenhoven}}]{qnoisetwolevel}
\bibinfo{author}{\bibfnamefont{R.}~\bibnamefont{Deblock}},
  \bibinfo{author}{\bibfnamefont{E.}~\bibnamefont{Onac}},
  \bibinfo{author}{\bibfnamefont{L.}~\bibnamefont{Gurevich}}, \bibnamefont{and}
  \bibinfo{author}{\bibfnamefont{L.~P.} \bibnamefont{Kouwenhoven}},
  \bibinfo{journal}{Science} \textbf{\bibinfo{volume}{301}},
  \bibinfo{pages}{203} (\bibinfo{year}{2003}).

\bibitem[{\citenamefont{Balestro et~al.}()\citenamefont{Balestro, Onac,
  Hartmann, Nazarov, and Kouwenhoven}}]{dotpat}
\bibinfo{author}{\bibfnamefont{F.}~\bibnamefont{Balestro}},
  \bibinfo{author}{\bibfnamefont{E.}~\bibnamefont{Onac}},
  \bibinfo{author}{\bibfnamefont{U.}~\bibnamefont{Hartmann}},
  \bibinfo{author}{\bibfnamefont{Y.~V.} \bibnamefont{Nazarov}},
  \bibnamefont{and} \bibinfo{author}{\bibfnamefont{L.~P.}
  \bibnamefont{Kouwenhoven}}, \emph{\bibinfo{title}{A quantum dot as a high
  frequency shot noise detector}}, \bibinfo{howpublished}{in preparation}.

\bibitem[{\citenamefont{Ingold and Nazarov}(1992)}]{IngoldNazarov}
\bibinfo{author}{\bibfnamefont{G.~L.} \bibnamefont{Ingold}} \bibnamefont{and}
  \bibinfo{author}{\bibfnamefont{Y.~V.} \bibnamefont{Nazarov}}, in
  \emph{\bibinfo{booktitle}{Single Charge Tunneling}}, edited by
  \bibinfo{editor}{\bibfnamefont{H.}~\bibnamefont{Grabert}} \bibnamefont{and}
  \bibinfo{editor}{\bibfnamefont{M.}~\bibnamefont{Devoret}}
  (\bibinfo{publisher}{Plenum Press}, \bibinfo{address}{New York},
  \bibinfo{year}{1992}), p.~\bibinfo{pages}{21}.

\bibitem[{\citenamefont{Kindermann et~al.}(2004)\citenamefont{Kindermann,
  Nazarov, and Beenakker}}]{kindermann:035336}
\bibinfo{author}{\bibfnamefont{M.}~\bibnamefont{Kindermann}},
  \bibinfo{author}{\bibfnamefont{Y.~V.} \bibnamefont{Nazarov}},
  \bibnamefont{and} \bibinfo{author}{\bibfnamefont{C.~W.~J.}
  \bibnamefont{Beenakker}}, \bibinfo{journal}{Phys. Rev. B}
  \textbf{\bibinfo{volume}{69}}, \bibinfo{eid}{035336} (\bibinfo{year}{2004}).

\bibitem[{\citenamefont{Rammer and Smith}(1986)}]{RammerSmith}
\bibinfo{author}{\bibfnamefont{J.}~\bibnamefont{Rammer}} \bibnamefont{and}
  \bibinfo{author}{\bibfnamefont{H.}~\bibnamefont{Smith}},
  \bibinfo{journal}{Rev. Mod. Phys.} \textbf{\bibinfo{volume}{58}},
  \bibinfo{pages}{323} (\bibinfo{year}{1986}).

\bibitem[{\citenamefont{Kindermann et~al.}(2003)\citenamefont{Kindermann,
  Nazarov, and Beenakker}}]{kindermcurvoltbias}
\bibinfo{author}{\bibfnamefont{M.}~\bibnamefont{Kindermann}},
  \bibinfo{author}{\bibfnamefont{Y.~V.} \bibnamefont{Nazarov}},
  \bibnamefont{and} \bibinfo{author}{\bibfnamefont{C.~W.~J.}
  \bibnamefont{Beenakker}}, \bibinfo{journal}{Phys. Rev. Lett.}
  \textbf{\bibinfo{volume}{90}}, \bibinfo{pages}{2468051}
  (\bibinfo{year}{2003}).

\bibitem[{\citenamefont{Kindermann and Nazarov}(2003)}]{KindermannNazarov2}
\bibinfo{author}{\bibfnamefont{M.}~\bibnamefont{Kindermann}} \bibnamefont{and}
  \bibinfo{author}{\bibfnamefont{Y.~V.} \bibnamefont{Nazarov}},
  \bibinfo{journal}{Phys. Rev. Lett.} \textbf{\bibinfo{volume}{91}},
  \bibinfo{pages}{136802} (\bibinfo{year}{2003}).

\end{thebibliography}
\end{document}